\documentclass[aps,twocolumn,showpacs,amsmath,amssymb,pra,superscriptaddress,floatfix,longbibliography]{revtex4-1}

\usepackage{times}
\usepackage{t1enc}
\usepackage{subfigure}
\usepackage{graphicx}
\usepackage{epstopdf}
\usepackage{amsmath}
\usepackage{dcolumn}
\usepackage{color} 
\usepackage{soul}

\begin{document}

% Use the \preprint command to place your local institutional report
% number in the upper righthand corner of the title page in preprint mode.
% Multiple \preprint commands are allowed.
% Use the 'preprintnumbers' class option to override journal defaults
% to display numbers if necessary
%\preprint{}

%Title of paper
\title{Photon induced atom recoil in collectively interacting planar arrays}

\author{Deepak A.~Suresh}
\affiliation{Department of Physics and Astronomy, Purdue University, West Lafayette,
Indiana 47907, USA}

\author{F.~Robicheaux}
\email{robichf@purdue.edu}
\affiliation{Department of Physics and Astronomy, Purdue University, West Lafayette,
Indiana 47907, USA}
\affiliation{Purdue Quantum Science and Engineering Institute, Purdue
University, West Lafayette, Indiana 47907, USA}

% repeat the \author .. \affiliation  etc. as needed
% \email, \thanks, \homepage, \altaffiliation all apply to the current
% author. Explanatory text should go in the []'s, actual e-mail
% address or url should go in the {}'s for \email and \homepage.
% Please use the appropriate macro foreach each type of information

% \affiliation command applies to all authors since the last
% \affiliation command. The \affiliation command should follow the
% other information
% \affiliation can be followed by \email, \homepage, \thanks as well.
%\author{}
%\email[]{Your e-mail address}
%\homepage[]{Your web page}
%\thanks{}
%\altaffiliation{}
%\affiliation{}

%Collaboration name if desired (requires use of superscriptaddress
%option in \documentclass). \noaffiliation is required (may also be
%used with the \author command).
%\collaboration can be followed by \email, \homepage, \thanks as well.
%\collaboration{}
%\noaffiliation

\date{April 30, 2021}

\begin{abstract}
The recoil of atoms in arrays due to the emission or absorption of photons is studied for  sub-wavelength interatomic spacing. The atoms in the array interact with each other through collective dipole-dipole interactions and with the incident laser field in the low intensity limit. Shining uniform light on the array gives rise to patterns of excitation and recoil in the array. These arise due to the interference of different eigenmodes of excitation. The relation between the recoil and the decay dynamics is studied when the array is in its excitation eigenstates. The recoil experienced by a subradiant collective decay is substantially larger than from independent atom decay. A method to calculate the rate of recoil when steady state has been achieved with a constant influx of photons is also described.

% insert abstract here
\end{abstract}

% insert suggested PACS numbers in braces on next line
\pacs{}
% insert suggested keywords - APS authors don't need to do this
%\keywords{}

%\maketitle must follow title, authors, abstract, \pacs, and \keywords
\maketitle

% body of paper here - Use proper section commands
% References should be done using the \cite, \ref, and \label commands
\section{Introduction\label{sec:intro}}

Collective dipole emissions have recently been used in many novel and interesting applications in the control of the interaction between light and matter. Reference \cite{dicke} showed that when an ensemble of atoms interact and radiate collectively, the dynamics are altered due to the interference of the outgoing light waves. This has led to abundant research investigating a wide variety of concepts including subradiance, superradiance, and collective Lamb shifts \cite{bettles,re1971,fhm1973,gfp1976,s2009,mss2014,pbj2014,bbl2016,bzb2016,psr2017,jbs2018}. These concepts are also being used in establishing a link between atoms separated by more than their resonant wavelength \cite{zoller,sbf2017,ggv2018}.

Placing the atoms in uniformly spaced arrays will enhance the co-operative response resulting in increased coupling between the atoms and radiation field \cite{bmp2017,wzc2015,bga2016,swl2017,chs2003,ymg2013,mmm2018,jrp2017,bpp2020}. Recent experiments have achieved the realisation of arrays of atoms with a high level of filling efficiency \cite{bds2016,ebk2016,llk2015,xlm2015,nlr2014,ghc2015,bloch}. Closely packed atom arrays, where the atom separation is less than the wavelength of the light, have been found to have highly reflective properties due to the cooperative dipole interactions \cite{bga2016,swl2017,bloch}. Such arrays have also been suggested to be efficient options for photon storage \cite{fjr2016,mma2018,ama2017}. 

Collective dipole interactions have found many applications in quantum information  \cite{zoller,mlx2015,wkw2016,xlm2015,ggv2018}. As such, quantum information processing requires an extremely high degree of fidelity and coherence. Many proposals ignore the effect of the recoil during the photon atom interactions. However, the recoil on the atoms can cause the information in the internal states to mix or entangle with the vibrational states of the atomic motion leading to a loss of coherence in the many atom electronic states. The subsequent motion can also affect the dynamics of the internal states of the system. Thus, the atom recoil can affect the overall robustness of the quantum system and introduce avenues for decoherence. The depth of the trapping potential required in experiments will be determined by this recoil and having a depth too low could escalate this problem.  But, the role of the photon recoil may be counter intuitive. For example,  the authors of Ref. \cite{bloch} discusses that their experiments showed that increasing the depth beyond a certain level resulted in a reduced cooperative response. This brings about the question of an optimal potential depth in experiments. Reference \cite{shahmoon} have also proposed to utilize atomic arrays to drive opto-mechanical systems and the recoil energy would play a large role in such interactions. Hence, it is imperative to accurately model the recoil force effects in such cases. 

The recoil in atoms have been studied in various other contexts. References \cite{cbp2009,gbd1992,jpl2004} describe how the atom recoil can alter the nature of interaction with light, followed by experimental observation of the same in Ref. \cite{wss2018}. Reference \cite{gag2018} have described a theoretical framework for the recoil from spontaneous emission in an atom near the interface of a topological medium. Reference \cite{amt2020} have also described an approach to analyze the recoil when light scatters from a Bose-Einstein condensate of a dilute gas with weak interatomic interaction. Reference \cite{ski2020} have shown that photon-induced recoil can induce quantum phase transitions in Waveguide Quantum Electrodynamic cold gas systems. Superradiance and the resulting atom recoil can induce a self-organization phase transitions in quantum gases \cite{bgb2010,mpdAr2021}.
The effect of quantized motion on the decay rates and shifts were calculated in Ref. \cite{dbm2016}.

The calculations in the current paper continue the exploration of the role of recoil in atomic ensembles conducted in Ref. \cite{shihua}. In the current paper, we study the recoil energy when photons interact with atomic ensembles, more specifically, ordered planar atomic arrays with sub-wavelength interatomic spacing. See Fig. \ref{fig:intro} for a schematic drawing. We describe a method to determine the recoil momentum and energy deposited in such cases using density matrix master equations. We focus on the regime where the periods of the atomic vibrations are much larger than the lifetimes of the internal excited states and the duration of the light pulse. We also work in the low light intensity regime, limiting the system to have one excited state at most, to reduce the computational complexity.

Reference \cite{shihua} treated simplified cases where the array size was considered near-infinite, ignoring the effects of the edges and assuming that most of the bulk of the array experiences a recoil to the same degree as the center atom. In the present work, we explore the role of finite array size and situations where the excitation of an atom strongly depends on its position in the array. Calculations for the individual atoms of the array show that the finiteness of the array has pronounced effects. We study the recoil and reflectance of the array when the atoms are driven by a pulsed excitation and a steady state excitation.

When all the atoms of the array are uniformly illuminated by light, interesting patterns arise in the recoil distribution as well as in the distribution of the excitation itself. This is due to the interference of the different eigenmodes of excitation, each associated with a modified lifetime and energy shift. We study these eigenmodes which leads to an understanding of the patterns, as well as the relation between the decay lifetimes and the recoil energies deposited in individual atoms. We also investigate the case where two nearly planar arrays act as a cavity, which have been proposed as interesting elements for connecting distant qubits. The prolonged lifetimes of cavities lead to enhanced recoil.

The paper is structured as follows. Section \ref{sec:methods} describes the methods used for the calculations. Section \ref{sec:results} presents results from a variety of situations. Section \ref{sec:conclusion} presents the conclusions followed by the Appendix.

\section{Methods\label{sec:methods}}
\begin{figure}
    \includegraphics[width=0.45\textwidth]{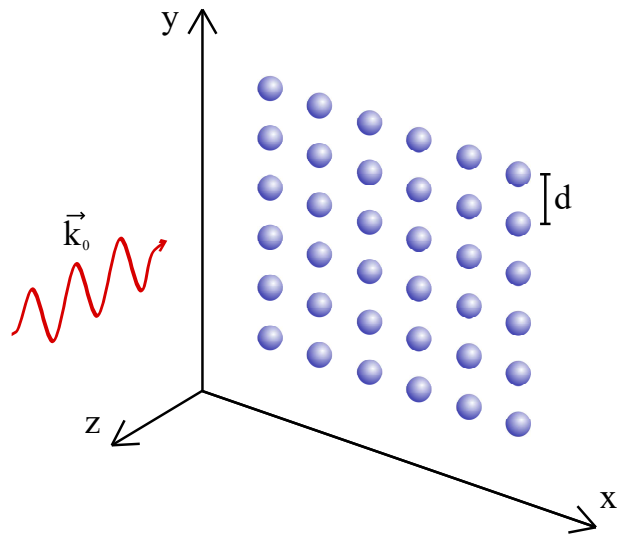}
    \caption{A schematic of the system considered. The atoms are the blue spheres and are placed in an ordered array in the x-y plane with interatomic separation $d$. The incoming light (red wavy arrow) is in the $\hat{\mathbf{z}}$ direction with wavevector $\mathbf{k}_0$}
    \label{fig:intro}
\end{figure}

The calculations are done for an ensemble of $N$ atoms arranged in a planar array with sub-wavelength interatomic spacing, $d$. The atoms have two internal energy levels which couple to each other through collective dipole-dipole interactions and with an external light field. The array lies in the x-y plane (with minor deviations in the z-direction when using curved arrays) and the direction of propagation of the incident laser light is chosen to be the z-direction ($\mathbf{k}_0 = k \hat{\mathbf{z}}$)  (Fig. \ref{fig:intro}). 
The light is circularly polarized in the $\hat{\mathbf{e}}_+ = -(\hat{\mathbf{x}} + i\hat{\mathbf{y}})/\sqrt{2}$ direction and the dipoles are also oriented in the $\hat{\mathbf{e}}_+$ direction. The circular polarization is chosen to form symmetric patterns in the array but the main conclusions should remain valid for other polarizations. Bold fonts denote vectors.

To describe the dynamics of the collective dipole emissions, we use the density matrix formalism where the density matrix evolves according to 
\begin{equation}
    \frac{d\hat{\rho}}{dt} = -\frac{i}{\hbar}[\hat{H}, \hat{\rho}] + \mathcal{L} (\hat{\rho})
    \label{eqn:rhodot}
\end{equation}
where $\hat{\rho}$ is the density matrix of the system, $\hat{H}$ is the effective Hamiltonian and $\mathcal{L}(\hat{\rho})$ is the Lindblad super-operator which describes the lossy collective dipole emissions. They shall be discussed subsequently.

To perform density matrix calculations for a large number of atoms, we adopt a few simplifications. We work in the low intensity limit where only the singly excited states are relevant, meaning only one atom can be excited at any time. In this limit, the density matrix dimensions scales more slowly with increasing number of atoms.  This simplification allows us to perform calculations up to $N = 300$ for the effects studied below. 

The ground state of the system which corresponds to all the atoms being in the electronic ground state, is represented by $|g\rangle= |g_1\rangle \otimes |g_2\rangle \otimes... |g_N\rangle$, where $|g_j\rangle$ denotes the electronic ground state of atom $j$. The raising and lowering operators of $j^{th}$ atom are represented  by $\hat{\sigma}^+_j$ and $\hat{\sigma}^-_j$ respectively. The state where only the $j^{th}$ atom is excited is represented by $|e_j\rangle = \hat{\sigma}^+_j |g\rangle$. 
The density matrix describes both the internal degrees of freedom and the positional dependence of the atoms. The positional dependence is based on the atoms' coordinates $\mathbf{r}_j$ and the density matrix is represented as 
\begin{equation}
\begin{split}
    \hat{\rho}\quad = \quad&\rho_{gg}(\mathbf{r}_1, \mathbf{r}_2, ...  \mathbf{r}_N;  \mathbf{r}_1', \mathbf{r}_2', ...  \mathbf{r}_N' ) |g\rangle \langle g|\\
    \sum_i &\rho_{e_ig}(\mathbf{r}_1, \mathbf{r}_2, ...  \mathbf{r}_N;  \mathbf{r}_1', \mathbf{r}_2', ...  \mathbf{r}_N' ) |e_i\rangle \langle g|\\
    \sum_{j}&\rho_{ge_j}(\mathbf{r}_1, \mathbf{r}_2, ...  \mathbf{r}_N;  \mathbf{r}_1', \mathbf{r}_2', ...  \mathbf{r}_N' ) |g\rangle \langle e_j|\\
    \sum_{i,j} &\rho_{e_ie_j}(\mathbf{r}_1, \mathbf{r}_2, ...  \mathbf{r}_N;  \mathbf{r}_1', \mathbf{r}_2', ...  \mathbf{r}_N' ) |e_i\rangle \langle e_j|\\
    \label{eqn:rho}
\end{split}
\end{equation}
where $i, j$ go from 1 to N. The coefficients describe the spatial dependence of the density matrix and are functions of $6N$ positional coordinates.

Since operators acting on the left and right sides of the density operator will act on different indices, we define the following convention
\begin{equation}
    \mathbf{r}_{ij} \equiv \mathbf{r}_i - \mathbf{r}_j;   \qquad      
    \mathbf{r}_{ij}' \equiv \mathbf{r}_i - \mathbf{r}_j';  \qquad    
    \mathbf{r}_{ij}'' \equiv \mathbf{r}_i' - \mathbf{r}_j'
\end{equation}

The effective Hamiltonian of the system will have (i) The laser interaction, (ii) the collective dipole interaction and (iii) the center of mass motional Hamiltonian consisting of the kinetic energy and the trapping potential of each atom. For the laser interaction, we work in the rotating wave approximation with 
\begin{equation}
    \hat{H}_L =  \hbar \sum_j \left[ -\delta\hat{\sigma}_j^+\hat{\sigma}_j^- + \frac{\Omega}{2}\hat{\sigma}_j^+ e^{i \mathbf{k}_0\cdot \mathbf{r}_j}+ \frac{\Omega^*}{2}\hat{\sigma}_j^- e^{-i \mathbf{k}_0\cdot \mathbf{r}_j}\right]
\end{equation}
which describes the Hamiltonian for a plane wave incident laser with $\Omega$ as the Rabi frequency, $\delta$ as the detuning and $\mathbf{k}_0 = k \mathbf{\hat{z}}$ as the initial wavevector of the incoming photons. The collective dipole emission also participates in the effective Hamiltonian based on the imaginary part of the Green's function ($Im\{g(\mathbf{r}_{ij})\}$) and is given by

\begin{equation}
    \hat{H}_{dd} = \hbar \sum_{i \neq j} Im\{g(\mathbf{r}_{ij}) \}\hat{\sigma}_i^+ \hat{\sigma}_j^-
\end{equation}
where the $g(\mathbf{r}_{ij})$ is the free space dyadic Green's function given by Eq. \eqref{eqn:greens}. When calculating the commutator of the effective Hamiltonian ($\hat{H} = \hat{H}_L + \hat{H}_{dd}$) with the density matrix $\hat{\rho}$ in Eq. \eqref{eqn:rhodot}, indices of the position vector must be carefully implemented. When right-multiplying $\hat{\rho}$ with $\hat{H}$, the corresponding primed indices must be used ($\mathbf{r}_j'$ for $\hat{H}_L$ and $\mathbf{r}_{ij}''$ for $\hat{H}_{dd}$).

The decay effects of the collective dipole interaction will be captured by the Lindblad term 
\begin{equation}
\begin{split}
    \mathcal{L}(\hat{\rho}) = \sum_{i,j}\big[&
    2Re\{g(\mathbf{r}_{ij}')\}\hat{\sigma}_i^- \hat{\rho} \hat{\sigma}_j^+  
    - Re\{g(\mathbf{r}_{ij})\}\hat{\sigma}_i^+ \hat{\sigma}_j^- \hat{\rho} \\
    &- Re\{g^*(\mathbf{r}_{ij}'')\}\hat{\rho}\hat{\sigma}_i^+ \hat{\sigma}_j^-
    \big]
\end{split}
\end{equation}
where the $i = j$ terms are the usual single atom decay part of the Lindblad operator. The Green's function $g(\mathbf{r}_{ij})$ is given by
\begin{equation}
\begin{split}
    g(\mathbf{r}_{ij}) = \frac{\Gamma}{2}&\bigg[
     \frac{3(\hat{r}_{ij}\cdot\hat{q}_i)(\hat{r}_{ij}\cdot\hat{q}_j^*) - (\hat{q}_i\cdot\hat{q}_j^*)}{2} h_2^{(1)}(kr_{ij})\\
     &+(\hat{q_i}\cdot\hat{q_j}^*)h_0^{(1)}( kr_{ij})
     \bigg]
     \label{eqn:greens}
\end{split}
\end{equation}
where $\hat{q}_i$ is the dipole orientation of the $i^{th}$ atom, $r_{ij} = |\mathbf{r}_{ij} |$ is the norm of $\mathbf{r}_{ij}$, $\hat{r}_{ij} = \mathbf{r}_{ij}/r_{ij}$ is the unit vector along $\mathbf{r}_{ij}$, $\Gamma$ is the decay rate of a single atom and $h_l^{(1)}$ are the outgoing spherical Hankel function of angular momentum $l$. When $i = j$, that is when $\mathbf{r}_{ij} = 0$, the imaginary part of the function becomes undefined, while the real part is defined. Hence, we redefine $g(\mathbf{r}_{ij})$ to be
\begin{equation}
\begin{aligned}
    g(\mathbf{r}_{ij}) &=  g(\mathbf{r}_{ij})&&for \quad i \neq j \\
    &=Re\{g(\mathbf{r}_{ij})\} &&for \quad  i = j
\end{aligned}
\end{equation}

\subsection{Slow Oscillation approximation\label{sec:slowoscillation}}

In this paper, the calculations are primarily in the limit where the timescales of the atomic motion in the trap are much slower than the timescales for the evolution of the internal states.  This means that during the evolution of the internal states, the position of the atom does not change. 
This also implies that the photon recoil can be considered as impulsive forces. This is a sudden approximation and allows us to drop the motional Hamiltonian with the kinetic energy and trapping potential. The recoil imparted by the photons and collective interactions can then be described by calculating the change in momentum ($\Delta\mathbf{p}$) and change in kinetic energy ($\Delta\mathbf{K}$) of the atoms. Typically, the frequencies of the trapping potentials will be around a few 10s of KHz and are much slower than the decay rates which are usually of the order of MHz. We also assume that the spread of the initial center of mass position is small compared to the separations. This will allow using the perfect position of the atoms in the calculations.

In the calculations where the system settles in the ground state, at infinite time,  only the $\rho_{gg}$ term is left non-zero and its spatial dependence will be of the form 

\begin{equation}
\begin{split}
    \rho_{gg} (\mathbf{r}_1, \mathbf{r}_2, ...;  \mathbf{r}_1', \mathbf{r}_2', ... ) = & \rho_0 (\mathbf{r}_1, \mathbf{r}_2, ...;  \mathbf{r}_1', \mathbf{r}_2', ... )\\
    &\times F (\mathbf{r}_1, \mathbf{r}_2, ...  ;  \mathbf{r}_1', \mathbf{r}_2', ... )
\end{split}
\end{equation}
where $\rho_0$ is the spatial dependence at the initial time and $F$ describes the evolution of the ground state spatial dependence. 
To calculate the momentum and kinetic energy change of the atoms, we take the expectation value of the corresponding operator using the density matrix.
\begin{equation}
\begin{split}
    \langle \mathbf{p}_j \rangle  & = \frac{\hbar}{i} \int [\hat{\nabla}_j\rho_{gg}] \delta(\mathbf{r}_{11}') \delta(\mathbf{r}_{22}')... d\mathbf{r}_1d\mathbf{r}_1'd\mathbf{r}_2d\mathbf{r}_2'...\\ 
    & = Tr[\hat{\mathbf{p}}_i (\rho_0 F)] = \langle \mathbf{p}_j \rangle_0 + \Delta \mathbf{p}_j
\end{split}
\end{equation}
where $\hat{\nabla}_j = \mathbf{\hat{x}} \frac{\partial}{\partial x_j} + \mathbf{\hat{y}} \frac{\partial}{\partial y_j}+ \mathbf{\hat{z}} \frac{\partial}{\partial z_j}$ is the momentum operator of the $j^{th}$ atom. The term $\langle \mathbf{p}_j \rangle_0$ denotes the initial expectation value of the momentum and derives from $\rho_0$. Hence, the change in momentum can be calculated from the function $F$ using
\begin{equation}
    \begin{split}
    \Delta \mathbf{p}_j   & = \frac{\hbar}{i} \int [\rho_0(\hat{\nabla}_j F)] \delta(\mathbf{r}_{11}') \delta(\mathbf{r}_{22}')... d\mathbf{r}_1d\mathbf{r}_1'd\mathbf{r}_2d\mathbf{r}_2'...\\ 
    & = Tr[\rho_0(\hat{\mathbf{p}}_i  F)]  
    \label{eqn:p}
\end{split}
\end{equation}

For the Kinetic energy, we follow a similar reasoning and use the KE operator $\mathbf{K}_j = \mathbf{p}_j^2/(2m)$ to get 
\begin{equation}
\begin{aligned}
    \Delta \mathbf{K}_j  & =  -\frac{\hbar^2}{2m} \int [\rho_0(\hat{\nabla}_j^2F)] \delta(\mathbf{r}_{11}') \delta(\mathbf{r}_{22}')... d\mathbf{r}_1d\mathbf{r}_1'd\mathbf{r}_2d\mathbf{r}_2'...\\
    & =  Re\{Tr[\rho_0(\hat{\mathbf{K}}_j F)]\}
    \label{eqn:KE}
\end{aligned}
\end{equation}

The density matrix is propagated in time using the Runge-Kutta second order integration until the system completely decays into the ground state. This ground state density matrix coefficient is used to evaluate $F(\mathbf{r}_1, \mathbf{r}_2, ...  ;  \mathbf{r}_1', \mathbf{r}_2', ... )$. This is in turn used to calculate the $\Delta \mathbf{p}$ and $\Delta \mathbf{K}$ in three dimensions using Eq. \eqref{eqn:p} and Eq. \eqref{eqn:KE} by using symmetric 2-point and 3-point differentiation.
To calculate the derivatives $\nabla_j$, the positions of either the primed ($\mathbf{r}_j$) or the unprimed coordinates ($\mathbf{r}_j'$) are shifted by a small distance $\delta r$ and evaluated.  For a more detailed explanation of the methods followed, refer to Section II of Ref. \cite{shihua}.

\subsection{Interatom distance\label{sec:interatomicdistance}}
The process of subradiance can be thought of as the destructive interference of the wavefunctions of the light emitted by spontaneous decay of excited atoms. For example, when there are two atoms close together to the point of overlap ($d \to 0$), the wavefunctions of the emitted light will cancel out if they have opposing phases. This results in a prolonged lifetime for the excited state. For two atoms, the out of phase state remains as the subradiant state until the separation of $d \sim \lambda /2$. As the separation goes beyond half a wavelength, the light acquires an extra phase of $\pi$ which causes the in-phase states to be subradiant until $d \sim \lambda$. This behavior continues as we increase $d$ and oscillates with a period of $\lambda/2$. The maximum effect of subradiance and superradiance possible decreases with increasing $d$ and becomes small beyond $d \sim \lambda$.

This behavior carries over to the periodically placed atoms in arrays.
For interatomic distances less than $\sim\lambda/2$, the subradiant states have adjacent atoms out-of-phase, while the superradiant states have them in-phase. Between $\lambda /2$ and $\lambda $, the subradiant states have adjacent atoms in-phase.
Since we want to primarily focus on subradiant states, and exciting atoms with adjacent atoms being in-phase is easier to experimentally realise, we choose the range of interatomic distance to be between $0.5 \lambda$ and $1.0 \lambda$.

% Put \label in argument of \section for cross-referencing
%\section{\label{}}
\section{Photon Recoil Energy and Momentum\label{sec:results}}
When a photon is absorbed or emitted by a single atom, the photon imparts a momentum kick $\hbar k$ and recoil kinetic energy $E_r = \hbar^2 k^2/(2m)$. But when there is an ensemble of atoms and collective dipole interactions take place, Ref. \cite{shihua} showed that the energy deposited is different and depends on collective decay dynamics. We delve deeper into this topic and discuss the directional properties of the kicks and its relation to the decay properties of the collective ensemble. Since a single photon undergoing perfect reflection on an atom imparts $2\hbar k$ momentum, the $\Delta p_z / (2\hbar k)$ serves as a good measure of reflectance.

The two different factors that contribute to the kicks have slightly different effects. In the out-of-plane direction, the collective dipole emissions emit light symmetrically on both sides and hence contribute to no net momentum kick in this direction, but will still contribute to the recoil energy deposited. In the in-plane direction, the atoms exchange photons among each other and, hence, the momentum and energy deposited will depend on the position of the atom within the array. The contribution from the laser will only be in the out of plane direction and has non-zero contributions to both momentum and kinetic energy deposition.

\subsection{Eigenstates\label{sec:eigenstates}}

Reference \cite{bettles} discussed the interference of many eigenmodes of the system which contributes to the cooperative emission. To get an understanding of the effect of the decay rate on the recoil energy, we analyze the photon recoil momentum and energy deposited when the initial state is an eigenstate of the excitation.
When there is no driving interaction with the laser, the eigenstate of the excitation is the eigenstate of the dyadic Green's function in free space. 
They are eigenstates of a complex symmetric matrix which means that they maintain the distribution pattern of the excitation among the atoms while the magnitude of the total excitation in the system decays exponentially.
We define the Green's tensor $G_{ij} = g(\mathbf{r}_{ij})$ which is an N$\times$N-dimensional matrix, and $\mathbf{V}_{\alpha}$ is an N-dimensional vector. The eigenvalue equation is 

\begin{equation}
    \sum_{j} G_{ij} \mathbf{V}_{j\alpha} = \mathcal{G}_{\alpha}\mathbf{V}_{i\alpha} = ( \frac{\gamma_{\alpha}}{2} + i \Delta_{\alpha})\mathbf{V}_{i\alpha}
\end{equation}
where $i$,$j$ are atom indices and $\alpha$ denotes the index for the eigenstate. $\mathcal{G}_\alpha$ is the eigenvalue and $\mathbf{V}_\alpha$ is the corresponding eigenvector. The rate of decay is given by the real part of the eigenvalue, $\gamma_{\alpha}$, and the shift in energy is given by the imaginary part, $\Delta_{\alpha}$. Since the Green's function is not Hermitian, the regular orthogonality conditions do not apply. Therefore, the vectors $\mathbf{V}_{\alpha}$ have to be normalized to satisfy,

\begin{equation}
    \sum_{i} \mathbf{V}_{i\alpha} \mathbf{V}_{i\alpha'} = \delta_{\alpha, \alpha'}
    \label{eqn:orthogonality}
\end{equation}
This relation should be used with care when there are degenerate eigenstates in the system. Each set of degenerate vectors must be orthogonalized to follow this condition.
These eigenstates form interesting and symmetric patterns on the array and exhibit similarities to TEM modes of light. As noted earlier in Section \ref{sec:interatomicdistance}, the adjacent atoms in the subradiant modes tend to be in-phase while the superradiant modes have the adjacent atoms out-of-phase in the specified range of $d$.

The array is initialized to one of the eigenstates, $\beta$, leading to the electronic part of the density matrix starting as $\rho(t=0) = \mathcal{N}_{\beta}|\mathbf{V}_{\beta}\rangle \langle \mathbf{V}_{\beta}|$, where the ket  $|\mathbf{V}_{\beta}\rangle = \sum_i \mathbf{V}_{i\beta} | e_i \rangle$ and 
$\mathcal{N}_{\beta} = (\sum_i |\mathbf{V}_{i\beta}|^2)^{-1}$ is a normalizing factor to have only one excitation at the initial time; note the magnitude in the definition of $\mathcal{N}_{\beta}$. The system is evolved in time until it completely reaches the ground state, at which point, the $\Delta \mathbf{p}$ and $\Delta \mathbf{K} $ of each atom are calculated.

The total kinetic energy deposited in the array for each eigenstate is inversely proportional to the decay rate of that state. This implies that highly subradiant states will have very high photon recoil energies. 
Another trend observed was that the kick on each individual atom was roughly proportional to the excitation probability of that atom. This, while being an expected result, allows for an easier way to look at the distribution of the energy deposition over the array. Commonly, the most subradiant mode at this range of interatomic distances is similar to a Gaussian distribution ($TEM_{00}$ like) on the array. This implies that the deposition of the recoil is concentrated near the center, while the atoms close to the edges have relatively low recoil. 

When the system is initialized to its eigenstates, it is possible to analytically obtain an expression to calculate the recoil after the system has decayed to the ground state. This expression can then be used to find the recoil in any arbitrary initial state configuration by decomposing it into its eigenstates. The coefficient of the ground state density matrix at infinite time will be 
\begin{equation}
    \rho_{gg}(\infty) = \sum_{\alpha, \alpha'}  \sum_{j,j'} 2 Re\{g(r_{jj'}')\} \mathbf{V}_{j\alpha} \mathbf{U}_{j'\alpha'}^*\frac{C_{\alpha,\alpha'}( 0)}{\mathcal{G}_{\alpha} + \mathcal{G}_{\alpha'}''^*}
    \label{eqn:rhoggEig}
\end{equation}
where $\mathcal{G}_{\alpha'}''$ and $\mathbf{U}_{j\alpha'}$ are the eigenvalue and eigenvector of the tensor $G_{ij}'' = g(\mathbf{r}_{ij}'')$. The contribution of each eigenstate is given by $C_{\alpha,\alpha'}(0) = \sum_{jj'} \mathbf{V}_{j\alpha} \mathbf{U}_{j'\alpha'}^* \rho_{e_je_{j'}}(0)$.
If the initial state is an eigenstate $\beta$, then it is given by $C_{\alpha,\alpha'}(0) = \mathcal{N}_\alpha \delta_{\alpha \beta}\sum_{i} \mathbf{V}_{i\beta}^* \mathbf{U}_{i\alpha'}^*$. and the term where $\alpha = \alpha' = \beta$ will be the most dominant. The derivation can be found in Appendix \ref{app:eigenstatedecay}. 

When initialized to eigenstates, this expression can be used to calculate the momentum and kinetic energy kicks using Eqs. \eqref{eqn:p} and \eqref{eqn:KE}. The calculation using Eq. \eqref{eqn:rhoggEig} is much easier than solving for large density matrices over extended times and it also provides intuition towards the trends observed. The denominator of the dominant term, $\mathcal{G}_{\beta} + \mathcal{G}_{\beta}''^* \approx 2 Re\{ \mathcal{G}_{\beta}\} = \gamma_{\beta}$ explains why the kicks are inversely proportional to the decay rates of the eigenstates. The term $\mathbf{V}_{j\beta} \mathbf{U}_{j'\beta}^*$ also explains why the kicks on each atom is proportional to their excitation probabilities.

An expression for kick in the out-of-plane direction can be obtained to express the $\Delta K_z$ depending only on the decay rate of the eigenvector ($\gamma_{\alpha}$). The derivation can be found in Appendix \ref{app:Kz}.

\begin{equation}
    \frac{\Delta K_z}{E_r} = \frac{2}{5} \frac{\Gamma}{\gamma _{\alpha}}
    \label{eqn:Kz}
\end{equation}

This was a surprising result because there is no photon mediated atom-atom interaction in the z-direction and there is just a single photon emitted. 
Hence one might expect the $\Delta K_z/E_r$ to be independent of the lifetime of the excited state. However, the results from Eq. \eqref{eqn:Kz} show that when there is spontaneous emission, the decay lifetimes play an important role in recoil of the atoms in the array. 

\begin{figure}
\centering
    \includegraphics[width=0.45\textwidth]{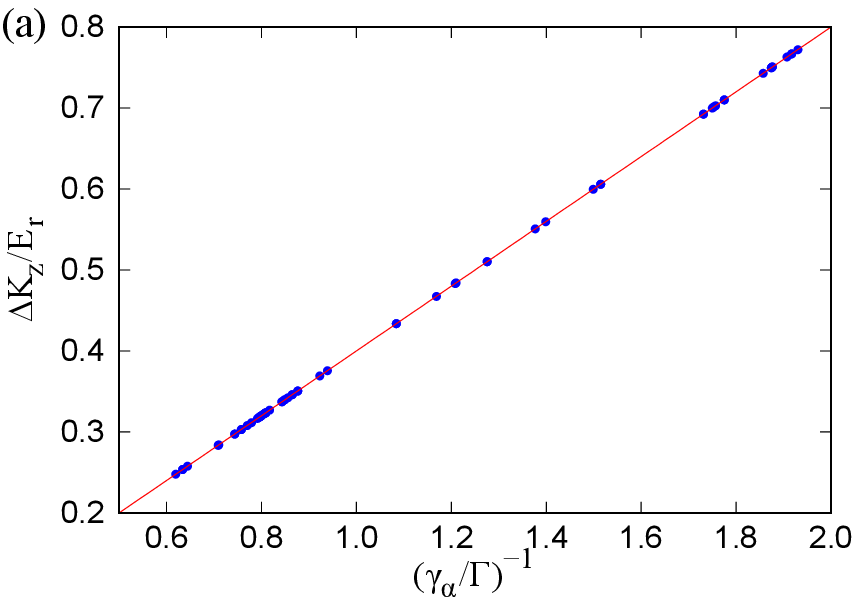}
    \includegraphics[width=0.45\textwidth]{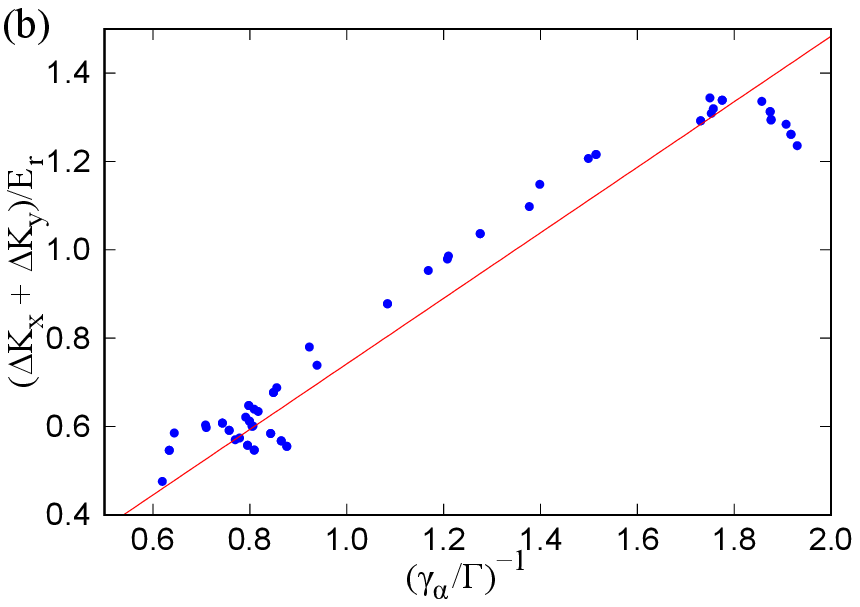}
\caption{The net kinetic energy kick and the linear fit for an array of $8\times8$ array, initially in an eigenstate, with respect to the inverse of the decay rate of the eigenvalue. Blue dots denote the calculated plot points and the red line denotes the linear fit of the data. (a) denotes the kick in the out-of-plane direction which is proportional to the lifetime with slope 2/5, as expected from Eq. \eqref{eqn:Kz}. (b) denotes the net kick in the plane of the array.}
\label{fig:Kxyz}
\end{figure}

Figure \ref{fig:Kxyz} shows the results for the net $\Delta \mathbf{K} $ deposited in the out-of-plane and in-plane  directions with respect to the decay lifetimes of the eigenstates using full density matrix calculations. While the out-of-plane kicks are highly proportional, the in-plane kicks have a more complicated dependence and are only roughly proportional.

This section discussed the recoil momentum and energy calculation for the case when the system is in either an eigenstate or an arbitrary excited state and allowed to decay into the ground state. This is a simplified case where there are no incoming photons and it is highly unlikely to be experimentally seen. Nevertheless, it provides insight into how the lifetimes of the state affects the recoil.

\subsection{Cavity\label{sec:cavity}}

The degree of subradiance can be further increased by trapping the light between two arrays and forming a cavity. By slightly curving the arrays, the cavity formed can greatly improve light retention, in turn making the system more subradiant. Reference \cite{zoller} have  proposed to utilize such curved mirrors to couple distant qubits, under the simplification of the atoms being fixed in space. This simplification may not be reasonable because the states are so subradiant that they give rise to enormous kicks on individual atoms. Hence in this section, we calculate the recoil experienced by the atoms in such highly subradiant cavities.

The system consists of $2N$ atoms making up two arrays each with $N$ atoms. The arrays are separated by a distance $L$. This is similar to the system describes in Fig. 1 of Ref. \cite{zoller}. To find the most efficient cavity mode, we find the eigenstates of the Green's function for the cavity and find the most subradiant mode. We vary the separation $L$ and the curvature of the array until we find the optimal parameters for maximum subradiance. The imaginary part of the eigenvalue also provides the detuning at which the cavity mode should be driven. 

The large lifetimes of the subradiant states prevent calculation of the kick in all the atoms of the array within reasonable computation time using the density matrix method. Hence we use the second method described in Sec. \ref{sec:eigenstates}  to calculate the recoil. 

To characterize the light retention capacity of the cavity, we calculate the finesse of the cavity. In this case, we define the finesse using the intensity distribution of the cavity. The cavity is driven using lasers of fixed detuning, while scanning the separation of the arrays. The finesse, $\mathcal{F}$, will be defined as the free spectral range, $\lambda / 2$, divided by the full width at half maximum in separation of the intensity:
\begin{equation}
    \mathcal{F} = \frac{\lambda/2}{\delta L_{FWHM}}
\end{equation}
This gives an estimate of the number of times a photon bounces in the cavity before decaying.

In the following sections, we discuss the results calculated for a few typical cavities with different decay rates. The cavities are initialized into their most subradiant eigenstate with a single excitation and evolved until they reach the ground state, similar to Sec. \ref{sec:eigenstates}.
The results are also compared with the finesse of the cavity. In each case, we take a $11 \times 11$  array and the separation has been optimized around $L = 20 \lambda$. 

\subsubsection{Spherical Mirrors\label{sec:cavitySpherical}}
\begin{figure}
    \centering
    \includegraphics[width=0.45\textwidth]{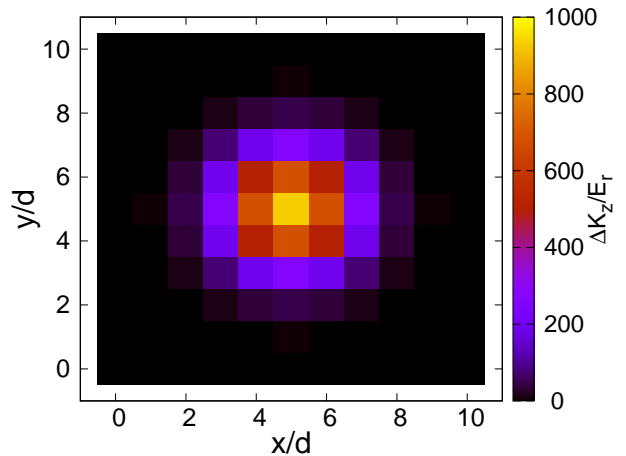}
    \caption{The recoil energy deposited in the z-direction on one array of a cavity with two $11\times11$  curved arrays with $d = 0.75 \lambda$ and $L = 19.75 \lambda$ corresponding to the system in Sec. \ref{sec:cavitySpherical}. Each cell denotes one atom in the array. Note the center atom has a kick of $\sim 900 E_r$.}
    \label{fig:cavity}
\end{figure}

Reference \cite{zoller} utilizes a set of spherically curved atom arrays to form highly subradiant states. 
Since light trapped in a cavity will have a Gaussian intensity profile, spherically curved mirrors match the spherical wavefront of the light and provides good energy retention. The mirrors are placed confocally. Following similar parameters to Ref. \cite{zoller}, with $d = 0.75 \lambda$, separation $L = 19.75 \lambda$, we attain the most subradiant mode to be extremely subradiant, with decay rates reaching $\sim 10^{-4}  \Gamma$. The finesse of this particular configuration was found to be 1250.
The kick the center atom experiences was calculated to be around $920 E_r$ in the out-of-plane direction and $33 E_r$ in the in-plane directions. This progressively reduces as we go closer to the edge in the shape of a Gaussian (See Fig. \ref{fig:cavity}). This trend is expected because the most subradiant eigenmode is typically a $TEM_{00}$ mode (as noted in the supplementary of Ref. \cite{zoller}). The atoms in the corner received a total of only $1.5\times 10^{-3} E_r$. The net kick on the array lead to an energy increase of $\sim 10364 E_r$.

Unfortunately, this result is not completely valid because the decay rate is too small for the slow oscillation approximation to be satisfied. If the duration of the force on the array is of the order of the oscillation period, the impulsive nature of the recoil force will not be valid. In this timescale, the exact nature of the way fields interact with the atoms is not precisely known. A fully quantum mechanical treatment considering the motion of the particles as quantum oscillators is required to get a better understanding of the dynamics. While such a calculation would be very difficult, the enormous size of the kicks from our simplified treatment suggest that an attempt should be made.

\subsubsection{Parabolic Mirrors}
Another typical type of mirrors used in cavities are parabolic mirrors. Confocal cavities are only marginally stable and any small errors in the positions of the mirrors will cause destabilization. Hence, we use a cavity in the region between confocal and concentric to provide some room for errors. This configuration, with $d = 0.8 \lambda$, parabolic focus $f = 15 \lambda$ and separation $L = 19.694 \lambda$, while not as subadiant as the previous case, has a decay rate of $\sim 10^{-3}  \Gamma$.  This decay rate also is at the edge of the slow oscillation approximation. The finesse for this cavity was calculated to be 263. The center atoms experienced $62 E_r$ recoil in the z-direction and $9.3 E_r$ in the x and y direction. A similar trend of decreasing kick in the edge atoms is seen and a total of only $8.3\times 10^{-4} E_r$ was deposited in the corner atom. The total energy deposited on the whole array in this decay process is $\sim 774 E_r$

\subsubsection{Plane mirrors}
Curving a plane array of atoms as required in the above cases may prove to be complicated. Hence, to compare, we discuss the case where a cavity is formed by two plane mirrors. In this configuration, with $d = 0.8 \lambda$ and separation $L = 20.045 \lambda$, the decay rates reach $\sim 10^{-2}  \Gamma$. The finesse was calculated to be approximately 14. The center atom received a total of $0.61 E_r$, while the corner atom received a total of $0.007 E_r$.
The array as a whole, received a kick of $10 E_r$ in the out-of-plane direction and $12 E_r$ in the in-plane directions.

\subsection{Pulsed Laser\label{sec:pulsed}}
To simulate the effects of a single photon interacting with the array, a low intensity pulse of light is incident on it. A planar array with $N$ atoms is initially in the ground state and a laser pulse with a Gaussian time profile is incident. The atoms are then evolved until they reach ground state. The $\Delta\mathbf{K}$ and $\Delta\mathbf{p}$ of each atoms are calculated in this time frame and are compared to the probabilities of excitation of each atom. The Gaussian time profile of the light pulse of the form
\begin{equation}
    \Omega(t) = \Omega_0  e^{-t^2/t_w^2}
\end{equation}
with peak Rabi frequency $\Omega_0$ and pulse width $t_w$. The $\Omega_0$ is kept low enough to not go beyond the single excitation limit. We see that like the trend in the previous section, the recoil is proportional to the integral of the excitation probabilities of each atom. That is, the excitation patterns are similar to the recoil distribution pattern in the array.

\begin{figure*}
        \centering
        \includegraphics[width=0.45\textwidth]{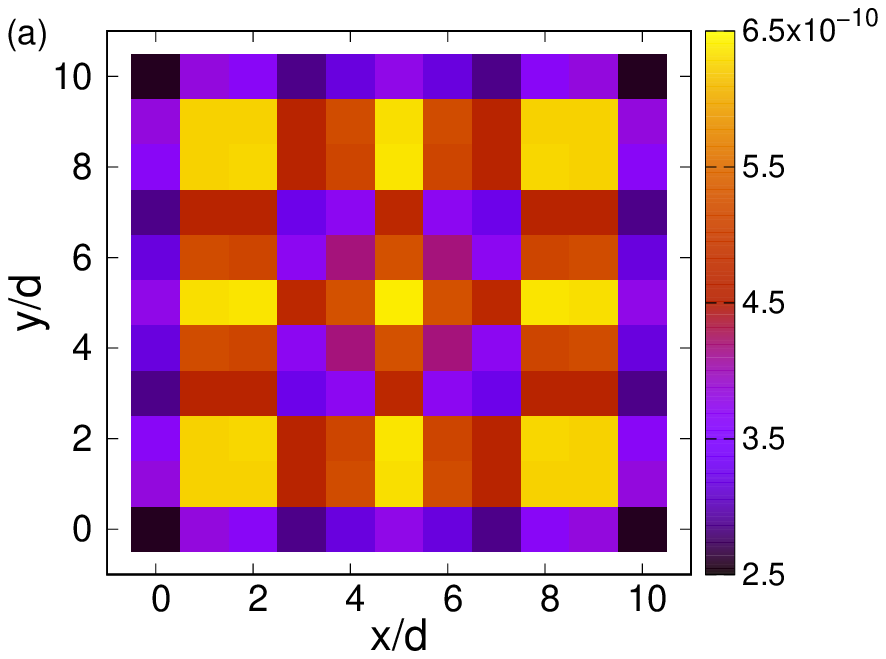}  
        \centering
        \includegraphics[width=0.45\textwidth]{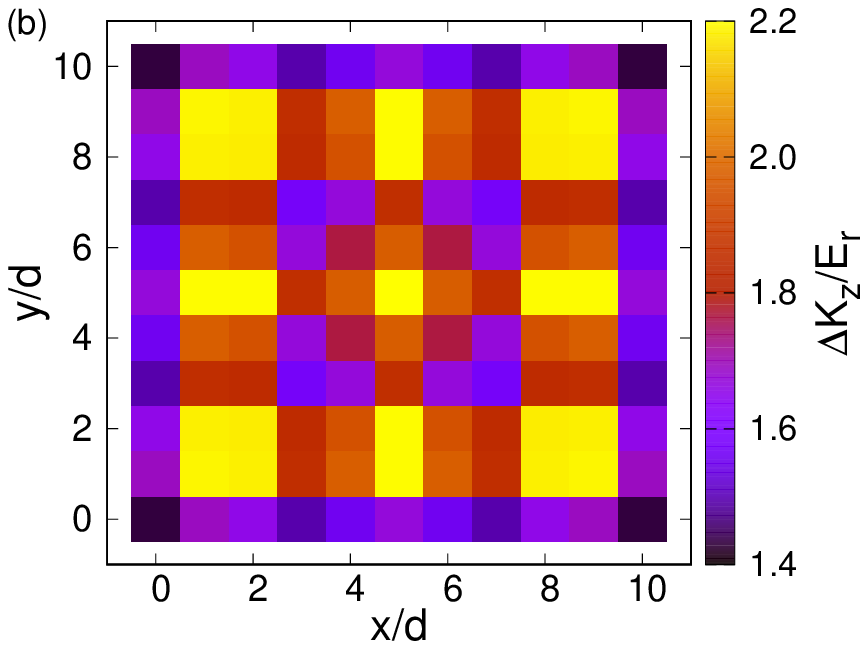}
    \caption{The excitation and recoil distribution pattern on a $ 11\times 11$ atom array with $d = 0.68 \lambda$ when excited by a laser pulse with peak Rabi frequency $\Omega_0 = 0.02 \Gamma$ and pulse width $t_w = 16/\Gamma$ at zero detuning. Each cell denotes one atom in the array. (a) shows the integral of the excitation probability in time for each atom in arbitrary units and (b) shows the $\Delta K_z/E_r$ deposited on each atom of the array per photon incident on the atom.  }
    \label{fig:ArEg}
\end{figure*}

This calculation is similar to in Ref. \cite{shihua}, except that Ref. \cite{shihua} assumes that there is a nearly uniform excitation of the atoms in the array and only calculates the dynamics of the central atom. By calculating the kicks in each atom of the array, we see patterns arise that vary with the detuning of the laser. Figure \ref{fig:ArEg} shows the comparison between the excitation pattern and the recoil distribution in the array. Figure \ref{fig:ArEg}(a) shows the time integrated excitation probability of each atom, while Fig. \ref{fig:ArEg}(b) depicts the  $\Delta K_z$ deposited on each atom. There is significant variation in the amount of energy deposited on the atoms and the corner atoms experienced only half the recoil energy of the atoms with the maximum recoil. These patterns are a combination of the eigenstates of the excitation as seen in Section \ref{sec:eigenstates}. The selection of the eigenstates depends on both the pattern as well as the detuning of the incoming light.

\begin{figure*}
    \centering
    \includegraphics[width=0.45\textwidth]{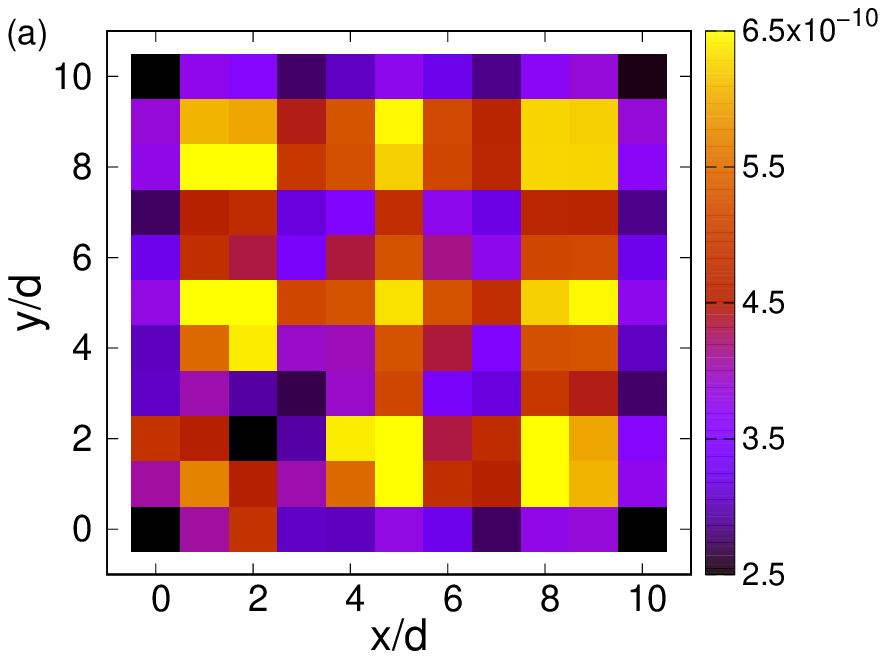}
    \includegraphics[width=0.45\textwidth]{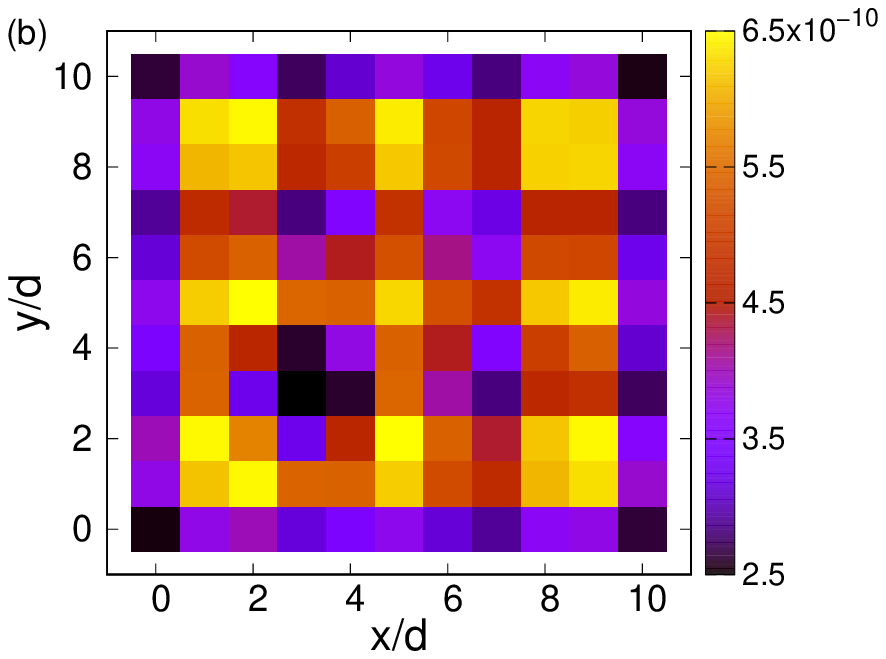}
    \caption{The effects of an atom missing from an array of $11\times11$ atoms with $d = 0.68 \lambda$ and pulsed light illumination with no detuning. The plots show the total integrated probability of excitation of each atom in arbitrary units. Compare with Fig. \ref{fig:ArEg}(a) which is the case where no atoms are missing. (a) has atom at coordinate (2,2) missing. It has a larger impact and affects up to second nearest neighbour atoms. (b) has atom at coordinate (3,3) missing. It has a smaller impact and only affects the immediate neighbours.}
    \label{fig:missing}
\end{figure*}

Experimentally, achieving a $100\%$ filling fraction for the array is difficult. Hence, the effects of the array missing a single atom have been studied. The scale of the disturbance caused by a missing atom depends on the contribution it would have made if present. If an atom is missing where the excitation is naturally weak, it has less effect on the excitation pattern and recoil (Fig. \ref{fig:missing}a). However, if the atom is missing where the excitation is large, there is a more substantial effect seen spanning a few nearby atoms (Fig. \ref{fig:missing}b). The recoil in the in-plane direction for the neighbouring atoms will also change. The recoil they experience in the direction towards the missing atom will be increased.

As explained in Section \ref{sec:interatomicdistance}, we have chosen the range of $d$ between $0.5 \lambda$ to $1.0 \lambda$ so that coupling subradiant modes, which have adjacent atoms in-phase, will be easier with uniform illumination. For other experiments that prioritize working with superradiant states, the interatomic distance could be chosen in the range $d < 0.5 \lambda$ for easy coupling with uniform light.

\subsection{Steady State\label{sec:steadystate}}

So far, we have studied the recoil received after the system has completely relaxed to the ground state at infinite time.
But in most experiments and applications, it is also important to understand the behavior of these kicks when there is a steady state involved. Instead of finding the total momentum and energy deposited, a more useful quantity will be the rate of energy deposited in each atom in steady state.

The methods described earlier as well as in Ref.  \cite{shihua} describe the process in which the final state of the system is the ground state. Hence we developed a different method to determine the rate of momentum and energy deposited. When the system has reached steady state, we can approximate that up to first order in time, all the density matrix elements except the ground state coefficient reaches equilibrium. This means that the only linear time dependent term will be the ground state density matrix coefficient ($\rho_{gg}$). Taking the positional derivatives on $\dot{\rho}_{gg}$ similar to Eq. \eqref{eqn:p} and Eq. \eqref{eqn:KE} yields the rate of the momentum and energy kicks. The density matrix can be evolved until steady state is reached and the $\dot{\rho}_{gg}$ can be calculated. Alternatively, the master equation can be solved analytically to obtain an expression for $\dot{\rho}_{gg}$.
\begin{equation}
\begin{split}
    \dot{\rho}_{gg}(t)  = &-i\sum_{j} \frac{\Omega^* (\mathbf{r}_j)}{2} \rho_{e_jg} +i\sum_{j} \frac{\Omega(\mathbf{r}_j')}{2} \rho_{ge_j}\\
    &+\sum_{ij} 2 Re\{g(\mathbf{r}_{ij}')\} \rho_{e_ie_j} 
\end{split}
\end{equation}
where $\Omega (\mathbf{r}_j)$ describes the spatial profile of the incoming laser as well as carrying a phase factor $ e^{-i\mathbf{k_0}\cdot\mathbf{r}_j}$.
The derivation and the procedure to calculate the density matrix terms at steady state are discussed in Appendix \ref{app:steadystate}. For this approximation to have good accuracy, it is important that we stay within the low intensity limit and use low Rabi frequencies.

The momentum and the energy deposited increase as a function of $\Omega^2$ which is expected as it is proportional to the number of incoming photons. To calculate the number of photons incident on an atom, we can integrate the intensity arriving on the area corresponding to one atom ($d^2$) per unit time and divide by the energy of a single photon ($\hbar c/\lambda_0$). The number of photons incident on each atom in one lifetime of a single atom is
\begin{equation}
    {\nu}_{_\Gamma} = \frac{2\pi}{3}  \left(\frac{d}{\lambda}\right)^2 \left(\frac{\Omega}{\Gamma}\right)^2
\end{equation}

The recent experimental work of Ref. \cite{bloch} to measure the reflectance of a subradiant atomic mirror with around $200$ atoms can be simulated.
For a $14\times14$ array at peak resonance, with $d = 0.68 \lambda$ and an influx of $1.8\times 10^{-4}$ photons per lattice site in one lifetime of a single atom, the calculations show that the average total energy deposited is $4.4 E_r$ per incoming photon. Each individual atom experienced a total recoil ranging from $2.6 E_r$ to $5.6 E_r$ per photon incident on each atom. The total average momentum imparted on each atom is $1.7 \hbar k$ per incoming photon, which corresponds to a reflectance of $85\%$.

Reference \cite{bga2016} discussed a null transmission situation for an infinite plane array with $d = 0.8 \lambda$. Figure 8 of Ref. \cite{shihua} also showed that the reflectance is near unity for an infinite array and estimates the same for a $11\times11$ array using the $\Delta p_z$ of the center atom. The emergence of the excitation patterns implies that this will be an overestimation for non-infinite arrays. 
Performing the recoil calculations on all the individual atoms shows that the reflectance only reaches $80\%$ for a $11\times11$ array when taking the average over all the atoms. 
The atoms at the edges couple less with the incoming light, decreasing the reflectance of the array. Re-calculating the average ignoring the edge atoms increases the reflectance to $95\%$.

By altering the detuning and the incident laser's transverse spatial profile, we can excite individual eigenmodes of the array. This would be impossible to experimentally implement for most states because of the high density of the atoms and the interatomic spacing being comparable to the resonant wavelength. The momentum imparted by the laser when exciting individual eigenmodes was was inversely proportional to the eigenmode’s decay rate. 
This is the result of the altered absorption rate of the incoming light. When the system is at steady state, the absorption rate must match the decay rate of that eigenmode. The energy deposited due to the collective dipole interaction followed a similar trend to the previous discussions and is also inversely proportional to the decay rate. This can be seen by the presence of the eigenvalue $\mathcal{G}_{\alpha}$ in the denominator of the dominant terms in the analytical solutions (Eq. \eqref{eqn:Adecompose}).

\begin{figure}
    \centering
    \includegraphics[width=0.45\textwidth]{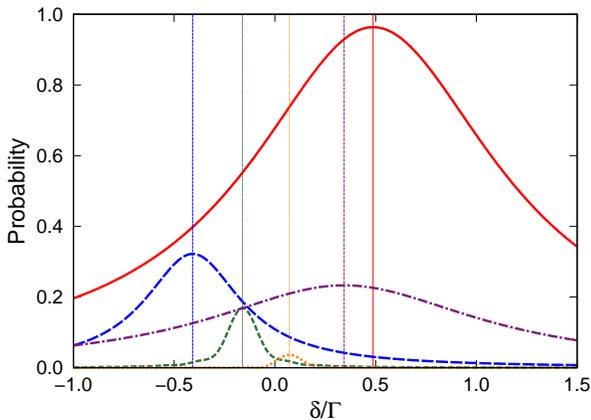}
    \caption{The decomposition of the state into its eigenmodes when excited by uniform light for a $5 \times 5$ array with $d = 0.4 \lambda$ versus the detuning $\delta$. The thick lines show the  probability of each eigenmode in rescaled arbitrary units while varying the detuning. The thin vertical lines show the line shift $\Delta_{\alpha}$ of the eigenmode with the corresponding color and dash-type. Only the 5 states that have a significant contribution are shown.
    }
    \label{fig:EigContribution}
\end{figure}

When driving with lasers, the detuning plays a role in deciding the contribution of the different eigenmodes as specified in Section \ref{sec:pulsed}. The eigenmodes closest to the symmetric Dicke state ($( \sum_i | e_i \rangle)/\sqrt{N}$) will couple better with the uniform incoming light.
Since each eigenstate is associated with an energy shift ($Im\{\mathcal{G}_\alpha\} = \Delta_\alpha$), matching the detuning also contributes to selecting the eigenstate. By using these factors, it is possible to influence which eigenstates contributes to the coupling of the atoms to the laser, to a certain extent. By using the orthogonality relations defined in Eq. \eqref{eqn:orthogonality}, we can determine the contribution from each eigenstate as we vary the detuning. As seen from Fig. \ref{fig:EigContribution}, the maximum contribution from a particular eigenstate is when the detuning matches with the shift associated with it. 

\subsection{Comparison to Trap Parameters}

There are various mechanisms used in experiments for trapping atoms, including optical lattices, tweezers, ion traps, etc. But in general, we can assume that the trap is deep enough to be approximated as a harmonic oscillator potential. The vibrational state of the atom in a quantum harmonic potential with trap frequency $\omega_{t}$ can be denoted by $n$. If the atom is initially in the ground state, the relation between the recoil energy deposited and the final expectation value of the vibrational state is given by
\begin{equation}
    \langle n \rangle = \frac{\Delta K}{\hbar \omega_{t}} 
\end{equation}

Typically, the trap frequencies used in experiments will be around $10$ kHz in the out-of-plane direction and $100$ kHz in the in-plane direction. For a Rubidium atom using the $780$ nm resonance, a recoil of $4 E_r$ due to the reflection of a single photon would bring the average vibrational state to $\langle n \rangle = 1.5$ and $0.15$ for $10$ kHz and $100$ kHz respectively.

If subradiance doubles the recoil, the average vibrational state reached also doubles. This causes unnecessary heating or loss of coherence in the system.  Even an $\langle n \rangle = 0.15$ could be an issue in quantum computing applications where the entanglement with the center of mass degrees of freedom must be less than $1\%$.
In the case of the calculation in Section \ref{sec:cavitySpherical}, the center atom receives $\sim 900 E_r$ in the decay process. Even with a high trap frequency of $100$ kHz, the center atom goes to the $\langle n \rangle = 35$ average vibrational state. The high vibrational state and the large spread over vibrational states would cause problems with decoherence in qubit implementations. The other atoms in the array would be in lower vibrational states and this uneven heating effect could lead to unforeseen issues.

\section{Conclusion\label{sec:conclusion}}

We presented the calculations for the recoil energy and momentum deposited in an array of atoms interacting collectively with light.
We studied the directional properties of the recoil and explored the effects of the eigenmodes of excitation in the system. The recoil energies added to each atom are proportional to the decay lifetimes of the eigenmodes and the more subradiant states experience more recoil than might be expected. This implies that recoil effects should be considered more carefully in experiments involving highly subradiant states. We note that when driving the array with uniform light, the excitation and recoil distribution in the array is not uniform but forms patterns which vary with the detuning. The patterns form due to the interference of the different eigenmodes coupling with the uniform light. These patterns also strongly contribute to the recoil of individual atoms. Calculations were done to determine the rate of energy deposition when the system reaches a steady state due to a constant influx of light. 

We also calculated the recoil effects in cavities made of curved atomic arrays and found pronounced effects as the decay rate decreased. However, our calculations are not applicable when the decay lifetimes become comparable to the timescale of the atomic vibration. Thus, our results for photon cavities (Sec. \ref{sec:cavity}) can be taken only as a qualitative estimate. It is unclear how the momentum transfer occurs on such timescales. A fully quantum approach would require large amounts of computational resources but exploring this situation with a few simplifying assumptions would give a better answer to this question. Reference \cite{shahmoon} explored the opposite regime where the internal state lifetimes are far larger than the motional timescale. It would be interesting to explore the physics at the interface of these two regimes. 

\begin{acknowledgments}
We thank Akilesh Venkatesh for critical reading an early version of this paper. This work was supported by the National Science Foundation under Award No. 1804026-PHY. This research was supported in part through computational resources provided by Information Technology at Purdue University, West Lafayette, Indiana.

\end{acknowledgments}

\appendix

\section{Eigenstate decay analytical calculation\label{app:eigenstatedecay}}

We derive an analytical method to calculate the kicks when the system starts off with a single excitation in any arbitrary excitation pattern. There is no driving from the laser interaction.   We write the state in terms of the eigenmodes $\mathbf{V}_{\alpha}$ and $\mathbf{U}_{\alpha}$ which are the eigenvectors of $G_{ij} = g(\mathbf{r}_{ij})$ and $G_{ij}'' = g(\mathbf{r}_{ij}'')$ respectively.
\begin{equation}
    \sum_{j'} G_{jj'} \mathbf{V}_{j'\alpha} = \mathcal{G}_{\alpha} \mathbf{V}_{j\alpha} 
    \quad
    \sum_{j'} G_{jj'}'' \mathbf{U}_{j'\alpha} = \mathcal{G}_{\alpha}'' \mathbf{U}_{j\alpha} 
    \label{eqn:AEigenvector}
\end{equation}
Since $G$ and $G''$ are non-Hermitian, their eigenvectors are not orthonormal and $\mathbf{V}_{\alpha}$ and $\mathbf{U}_{\alpha}$ have to be normalized to follow
\begin{equation}
    \sum_{j} \mathbf{V}_{j\alpha} \mathbf{V}_{j\alpha'} = \delta_{\alpha\alpha'}
    \quad
    \sum_{j} \mathbf{U}_{j\alpha} \mathbf{U}_{j\alpha'} = \delta_{\alpha\alpha'}
    \label{eqn:AOrtho}
\end{equation}
Hence, we can write the $g(\mathbf{r}_{ij}$)s as follows
\begin{equation}
\begin{split}
    g(\mathbf{r}_{ij}) = \sum_{\alpha}\mathbf{V}_{i\alpha}\mathbf{V}_{j\alpha}\mathcal{G}_{\alpha}
    \\
    g^*(\mathbf{r}_{ij}'') = \sum_{\alpha}\mathbf{U}_{i\alpha}^*\mathbf{U}_{j\alpha}^*\mathcal{G}_{\alpha}^*
\end{split}
\end{equation}
In the case where there is no laser interaction, the terms $\rho_{e_jg}$ and $\rho_{ge_j}$ will vanish giving
\begin{equation}
    \hat{\rho}(t) = \rho_{gg}(t)|g\rangle \langle g| +\sum_{jj'} \rho_{e_je_{j'}}(t) \hat{\sigma}_j^+ |g\rangle \langle g| \hat{\sigma}_{j'}^-  
\end{equation}
where the positional dependence has been suppressed for notational convenience. We can write the density matrix equation as 
\begin{equation}
\begin{split}
      \frac{d\hat{\rho}}{dt} = \sum_{i,j}\big[&
    - g(\mathbf{r}_{ij})\hat{\sigma}_i^+ \hat{\sigma}_j^- \hat{\rho} 
    - g^*(\mathbf{r}_{ij}'')\hat{\rho}\hat{\sigma}_i^+ \hat{\sigma}_j^-\\
    &+2Re\{g(\mathbf{r}_{ij}')\}\hat{\sigma}_j^- \hat{\rho} \hat{\sigma}_i^+ 
    \big]
    \label{eqn:Arhodot}
\end{split}
\end{equation}
Solving this we get,
\begin{equation}
    \frac{d\rho_{e_je_{j'}}(t)}{dt} = \sum_k \left( -g(\mathbf{r}_{jk})\rho_{e_ke_{j'}} -\rho_{e_je_k}g^*(\mathbf{r}_{kj'}'') \right)
\end{equation}
Decomposing the $\rho_{e_je_{j'}}$ in terms of the eigenfunctions $\mathbf{V}_{\alpha}$ and $\mathbf{U}_{\alpha'}^*$
\begin{equation}
    \rho_{e_je_{j'}} = \sum_{\alpha \alpha'} \mathbf{V}_{j\alpha}\mathbf{U}_{j'\alpha'}^* C_{\alpha \alpha'}
    \label{eqn:ACjj'}
\end{equation}
Using this, we get
\begin{equation}
\begin{split}
    \frac{dC_{\alpha\alpha'}}{dt} = - (\mathcal{G}_{\alpha} +\mathcal{G}_{\alpha'}''^* ) C_{\alpha\alpha'}\\
       \implies C_{\alpha\alpha'}(t) = C_{\alpha\alpha'}(0) e^{- (\mathcal{G}_{\alpha} +\mathcal{G}_{\alpha'}''^* )t}
\end{split}
\end{equation}

Since the $\rho_{gg}$ term only arises from the last term of Eq. \eqref{eqn:Arhodot}, we get
\begin{equation}
\begin{split}
    \frac{d\rho_{gg}}{dt} &= \sum_{jj'} 2Re\{g(\mathbf{r}_{jj'})\} \rho_{e_je_{j'}}\\
    &= \sum_{\alpha\alpha'}  \sum_{jj'} 2Re\{g(\mathbf{r}_{jj'})\}\mathbf{V}_{j\alpha}\mathbf{U}_{j'\alpha'}^* C_{\alpha \alpha'}(t)
\end{split}
\end{equation}

Since $C_{\alpha \alpha'}(t) $ are exponentials, 
\begin{equation}
    \rho_{gg}(\infty) = \sum_{\alpha, \alpha'}  \sum_{j,j'} 2 Re\{g(r_{jj'}')\} \mathbf{V}_{j\alpha} \mathbf{U}_{j'\alpha'}^*\frac{C_{\alpha,\alpha'}( 0)}{\mathcal{G}_{\alpha} + \mathcal{G}_{\alpha'}''^*}
    \label{eqn:Arhogg}
\end{equation}
The expression for $C_{\alpha\alpha'}(0)$ can be obtained from Eq. \eqref{eqn:ACjj'} using the orthogonality relations (Eq. \eqref{eqn:AOrtho}) to be
\begin{equation}
    C_{\alpha,\alpha'} = \sum_{jj'} \mathbf{V}_{j\alpha} \mathbf{U}_{j'\alpha'}^* \rho_{e_je_{j'}}
\end{equation}
If the system is initialized to one of its eigenstates $\beta$ with $\rho(t=0) = \mathcal{N}_{\beta}|\mathbf{V}_{\beta}\rangle \langle \mathbf{V}_{\beta}|$, then
\begin{equation}
    C_{\alpha,\alpha'}(t=0) = \mathcal{N}_{\beta} \delta_{\alpha \beta}\sum_{i} \mathbf{V}_{i\beta}^* \mathbf{U}_{i\alpha'}^*
    \label{eqn:ACalphaalpha'}
\end{equation}
where  $\mathcal{N}_{\beta} = (\sum_i |\mathbf{V}_{i\beta}|^2)^{-1}$ is the normalization factor.

\subsection{Out-of-plane eigenstate recoil energy\label{app:Kz}}
When we analyze the kinetic energy kick imparted on the out-of-plane direction when the system is initialized to an eigenstate, we get a proportionality equation between the energy deposited and the state lifetimes. This can be derived by taking the second derivative from Eq. \eqref{eqn:KE} only in the z-direction. When the system is initialized to an eigenstate $\beta$, the dominant terms in the summation of Eq. \eqref{eqn:Arhogg} will be the terms with $\alpha = \alpha' = \beta$. Hence the summation over the eigenstates will vanish giving
\begin{equation}
    \rho_{gg}(\infty) =  \sum_{j,j'} 2 Re\{g(r_{jj'}')\} \mathbf{V}_{j\beta} \mathbf{U}_{j'\beta}^*\frac{C_{\beta,\beta}( 0)}{\mathcal{G}_{\beta} + \mathcal{G}_{\beta}''^*}
\end{equation}

To take derivatives in the z-direction, we shift the primed coordinates, $\mathbf{r}_j' = \mathbf{r}_j + \delta z$. This would result an insignificant change in $\mathcal{G}_{\beta}'' = \mathcal{G}_{\beta} + 10^{-2} O(\delta z^3)$ and a second order change in the eigenvectors, $\mathbf{U}_{j\beta} = \mathbf{V}_{j\beta} + 10^{-2} O(\delta z^2)$. Hence when taking up to second derivatives, we can assume $\mathcal{G}_{\beta}'' \approx \mathcal{G}_{\beta}$ and $\mathbf{U}_{j\beta} \approx \mathbf{V}_{j\beta}$. This implies that $\mathcal{G}_{\beta} + \mathcal{G}_{\beta}''^* = 2 Re\{\mathcal{G}_{\beta}\} = \gamma_{\beta}$ and $\mathbf{V}_{j\beta} \mathbf{U}_{j\beta}^* = |\mathbf{V}_{j\beta}|^2$. Using Eq. \eqref{eqn:ACalphaalpha'},  $C_{\beta,\beta}(0)$ will also become $\mathcal{N}_{\beta}$. This gives

\begin{equation}
     \rho_{gg}(\infty) =  \sum_{j,j'} 2 Re\{g(r_{jj'}')\} \mathbf{V}_{j\beta}\mathbf{U}_{j'\beta}^* \frac{\mathcal{N}_{\beta}}{\gamma_{\beta}}
\end{equation}

When taking the second derivative for the atom $k$, only the term $j = j' = k$ will be non vanishing. 

\begin{equation}
     \rho_{gg}(\infty) =   2 Re\{g(r_{kk}')\} |\mathbf{V}_{k\beta}|^2 \frac{\mathcal{N}_{\beta}}{\gamma_{\beta}}
\end{equation}

At the limit of $\delta z \to 0$, 
\begin{equation}
    \lim_{\delta z \to 0} \frac{\partial^2 Re\{g(\mathbf{r}_{kk}')\}}{\partial z ^2} = -k^2\frac{\Gamma}{5}
\end{equation}

Hence the kinetic energy deposited on atom $k$ will be
\begin{equation}
    \Delta K_z^{(k)} = \frac{\hbar^2 k^2}{2m}\frac{2\Gamma}{5} |\mathbf{V}_{k\beta}|^2 \frac{\mathcal{N}_{\beta}}{\gamma_{\beta}} 
\end{equation}
The total kinetic energy deposited on the whole array in the z-direction will be (using $\mathcal{N}_{\beta} = (\sum_i |\mathbf{V}_{i\beta}|^2)^{-1}$)

\begin{equation}
    \frac{\Delta K_z}{E_r} = \frac{2}{5} \frac{\Gamma}{\gamma_{\beta}}
\end{equation}

\section{Steady state analytical calculation\label{app:steadystate}}

In this appendix, we derive analytically a method to find the recoil at steady state by decomposing the state into its eigenstates. For simplicity in this derivation, $\rho_{e_jg}$, $\rho_{ge_j}$ and $\rho_{e_ie_j}$ will be represented as $w_j$, $\tilde{w}_j$ and $\tilde{\rho}_{ij}$ respectively and the positional dependence will not be explicitly shown. That is, Eq. \eqref{eqn:rho} will be represented as

\begin{equation}
\begin{split}
    \hat{\rho}\quad = \quad&\rho_{gg} |g\rangle \langle g|
    +\sum_i w_i |e_i\rangle \langle g|\\
    +&\sum_{j}\tilde{w}_j |g\rangle \langle e_j|
    +\sum_{i,j} \tilde{\rho}_{ij} |e_i\rangle \langle e_j|\\
\end{split}
\end{equation}

A constant laser with detuning $\delta$ and Rabi frequency $\Omega(\mathbf{r}_j)$ is incident on the $j^{th}$ atom of the array. By using Eq. \eqref{eqn:rhodot}, we can obtain the rate of change of the coefficients of the density matrix. 

\begin{subequations}

\begin{equation}
    \dot{w}_{j} = -i \frac{\Omega(\mathbf{r}_j)}{2} \rho_{gg} + i\delta w_{j} - \sum_k g(\mathbf{r}_{jk})  w_{k}
    \label{eqn:Acoeff1}
\end{equation}
\begin{equation}
    \dot{\tilde{w}}_{j} = i \frac{\Omega^*(\mathbf{r}_j')}{2} \rho_{gg} - i\delta \tilde{w}_{j} - \sum_k g^*(\mathbf{r}_{jk}'')  \tilde{w}_{k}
    \label{eqn:Acoeff2}
\end{equation}

\begin{equation}
\begin{split}
    \dot{\tilde{\rho}}_{ij} = &-i \frac{\Omega(\mathbf{r}_i)}{2} \tilde{w}_j + i \frac{\Omega^*(\mathbf{r}_j')}{2} w_i\\
    &-\sum_k g(\mathbf{r}_{ik}) \tilde{\rho}_{kj}
    -\sum_k g^*(\mathbf{r}_{kj}'') \tilde{\rho}_{ik}
    \label{eqn:Acoeff3}
\end{split}
\end{equation}

\begin{equation}
\begin{split}
    \dot{\rho}_{gg} = &\sum_{ij} 2 Re\{g(\mathbf{r}_{ij}')\} \tilde{\rho}_{ij} \\
    -i&\sum_j \frac{\Omega^*(\mathbf{r}_j)}{2} w_j
    +i\sum_j \frac{\Omega(\mathbf{r}_j')}{2} \tilde{w}_j
    \label{eqn:Acoeff4}
\end{split}    
\end{equation}

\end{subequations}

where $\Omega (\mathbf{r}_j)$ describes the spatial profile of the incoming laser as well as carrying a phase factor $ e^{-i\mathbf{k_0}\cdot\mathbf{r}_j}$.
We can decompose the coefficients $w_j$, $\tilde{w}_j$ and $\tilde{\rho}_{ij}$ in terms of the eigenstates $\mathbf{V}_{j\alpha}$ and $\mathbf{U}_{j\alpha}$ (defined in Eq. \eqref{eqn:AEigenvector}) to get $w_{\alpha}'$, $\tilde{w}_{\alpha}'$ and $\tilde{\rho}_{\alpha\alpha'}'$ using

\begin{subequations}
\begin{equation}
    w_{\alpha}' = \sum_j \mathbf{V}_{j\alpha} w_j\qquad
    w_j = \sum_{\alpha} \mathbf{V}_{j\alpha} w_{\alpha}'
    \label{eqn:Adecompose1}
\end{equation}

\begin{equation}
    \tilde{w}_{\alpha}' = \sum_j \mathbf{U}^*_{j\alpha} \tilde{w}_j\qquad
    \tilde{w}_j = \sum_{\alpha} \mathbf{U}^*_{j\alpha} \tilde{w}_{\alpha}'
    \label{eqn:Adecompose2}
\end{equation}

\begin{equation}
    \tilde{\rho}_{\alpha\alpha'}' = \sum_{ij} \mathbf{V}_{i\alpha} \mathbf{U}^*_{j\alpha} \tilde{\rho}_{ij}\qquad
    \tilde{\rho}_{ij} = \sum_{\alpha\alpha'} \mathbf{V}_{i\alpha} \mathbf{U}^*_{j\alpha} \tilde{\rho}_{\alpha\alpha'}'
    \label{eqn:Adecompose3}
\end{equation}
\label{eqn:Adecompose}
\end{subequations}

In the first order in time, the recoil will only build up on the ground state and the change in the other coefficients will be zero, i.e., $\dot{w}_{j}$, $ \dot{\tilde{w}}_{j}$ and $\dot{\tilde{\rho}}_{ij}$ will  be zero. Therefore, we can solve for $w_{\alpha}'$, $\tilde{w}_{\alpha}'$ and $\tilde{\rho}_{\alpha\alpha'}'$ using Eqs. \eqref{eqn:Acoeff1}, \eqref{eqn:Acoeff2} and \eqref{eqn:Acoeff3},  which gives

\begin{subequations}

\begin{equation}
    w_{\alpha}' = \frac{-i \rho_{gg}}{2} \sum_i \frac{\Omega(\mathbf{r}_i)\mathbf{V}_{i\alpha}}{\mathcal{G}_{\alpha} -i\delta}\\
\end{equation}
\begin{equation}
    \tilde{w}_{\alpha}' = \frac{i \rho_{gg}}{2} \sum_i \frac{\Omega^*(\mathbf{r}_i')\mathbf{U}^{*}_{i\alpha}}{\mathcal{G}_{\alpha}''^{*} +i\delta}\\
\end{equation}
\begin{equation}
    \tilde{\rho}_{\alpha\alpha'}' = \frac{\rho_{gg}}{4}
    \sum_i \frac{\Omega(\mathbf{r}_i)\mathbf{V}_{i\alpha}}{\mathcal{G}_{\alpha} -i\delta}
    \sum_j \frac{\Omega^*(\mathbf{r}_j')\mathbf{U}^*_{j\alpha'}}{\mathcal{G}_{\alpha'}''^{*} +i\delta}
\end{equation}
\end{subequations}

The term $\rho_{gg}$ can be calculated using $Tr[\hat{\rho}] = 1$, but in the low intensity limit, it will be close to $1$. These equations can be used in conjunction with Eqs. \eqref{eqn:Acoeff4} and \eqref{eqn:Adecompose} to calculate $\dot{\rho}_{gg}$ to find the recoil kinetic energy and momentum.

\bibliography{ref.bib}

\end{document}